\documentclass[aps,prl,floatfix,twocolumn,showpacs,10pt]{revtex4-1}

\usepackage{graphicx}
\usepackage{dcolumn}
\usepackage{bm}
\usepackage{amssymb}
\usepackage{rotating} 
\usepackage{xcolor}

\begin{document}

\title{Exact Two-body Expansion of the Many-particle Wave Function}

\author{David A. Mazziotti}

\email{damazz@uchicago.edu}
\affiliation{Department of Chemistry and The James Franck Institute, The University of Chicago, Chicago, IL 60637}%

\date{Submitted May 20, 2020{\textcolor{black}{; Revised August 27, 2020}}}


\begin{abstract}

 Progress toward the solution of the strongly correlated electron problem has been stymied by the exponential complexity of the wave function.  \textcolor{black}{Previous work established an exact two-body exponential product expansion for the ground-state wave function.  By developing a reduced density matrix analogue of Dalgarno-Lewis perturbation theory, we prove here that (i) the two-body exponential product expansion is rapidly and globally convergent with each operator representing an order of a renormalized perturbation theory,  (ii) the energy of the expansion converges quadratically near the solution, and (iii) the expansion is exact for both ground and excited states.  The two-body expansion offers a reduced parametrization of the many-particle wave function as well as the two-particle reduced density matrix with potential applications on both conventional and quantum computers for the study of strongly correlated quantum systems.}  We demonstrate the result with the exact solution of the contracted Schr{\"o}dinger equation for the molecular chains H$_{4}$ \textcolor{black}{and H$_{5}$}.

\end{abstract}


\maketitle

{\em Introduction:} Computations of strongly correlated electrons are critical to the study of molecules and the prediction of their properties  from medicine to materials.  However, such computations are stymied by the exponential complexity of the wave function.  Significant progress has been made in two general directions: (1) circumventing the wave function by expressing the energy with few electron quantities like the two-electron reduced density matrix~\cite{Coleman2000, Mazziotti2007, Nakata2001, Mazziotti2002b, Mazziotti2004a, Zhao2004, Cances2006, Erdahl2007, Shenvi2010, Mazziotti2011, Mazziotti2012b, Verstichel2012, Poelmans2015, Mazziotti2016, Schlimgen2016, Mostafanejad2018, Xie2020, Mazziotti2006, Mazziotti2007a, Snyder2011c, Sand2015, Smart2018} (or even the 1-electron reduced density matrix~\cite{Piris2007, Lathiotakis2009, Piris2017a, Schmidt2019, Schilling2019}) and (2) simplifying the structure of the wave function in theories like coupled cluster~\cite{Bartlett2007,Pawlowski2019}, density matrix renormalization group~\cite{Schollwoeck2005, Chan2011}, and stochastic or sparse configuration interaction methods~\cite{Booth2009, Li2018, Xu2018, Wang2019}.  In this Communication we show that the pairwise nature of the electron interactions that is central to reduced density matrix methods can be exploited to further simplify the structure of the wave function.

While two-body expansions of the wave function have been previously proposed~\cite{Nakatsuji2000, Nooijen2000, Voorhis2001, Nakatsuji2001, Davidson2003, Ronen2003, Piecuch2003, Mazziotti2004c, Kutzelnigg2005} and implemented~\cite{Mazziotti2006, Mazziotti2007a, Snyder2011c, Sand2015, Smart2018, Yanai2006, Li2016, Maitra2017}, questions about their rates of convergence and their applicability to arbitrary stationary states have not been adequately addressed.  \textcolor{black}{It was originally conjectured that a single exponential of a general two-body operator could produce the exact ground-state wave function from a Slater determinant reference~\cite{Nooijen2000, Voorhis2001, Davidson2003, Ronen2003, Piecuch2003, Mazziotti2004c, Kutzelnigg2005}.   While this conjecture is false, we showed in previous work that an exact ground-state wave function with a size-extensive energy can be generated from a product of exponentials of two-body operators on a Slater determinant reference~\cite{Mazziotti2004c}.}    Recently, there has been renewed interest in such expansions, especially in the context of molecular simulations on near-term quantum computers~\cite{Lee2018, Grimsley2019, Evangelista2019, Smart2020}.  \textcolor{black}{Here we extend our previous work to show that the two-body exponential-product expansion is rapidly convergent and exact for both ground and excited states.  By developing a contraction of Dalgarno-Lewis perturbation theory~\cite{1955a, Schwartz1959, Schiff1968} onto the space of two particles, we prove that, although a single exponential of a two-body operator is not exact in general, it is sufficient to generate any wave function within the reach of a renormalized first-order perturbation theory about the reference wave function.  Consequently, the product of two-body exponential operators can move the wave function efficiently through the ground or excited states of a family of Hamiltonians connecting the reference state to the target state.}  For any stationary state the expansion is globally convergent with local quadratic convergence in the energy.   We demonstrate the theory with the exact solution of the contracted Schr{\"o}dinger equation~\cite{Nakatsuji1976, Colmenero1993a, Nakatsuji1996, Yasuda1997, Mazziotti1998b, Mazziotti1999a, Coleman2000, Mazziotti2002d} for \textcolor{black}{ground and excited states of} the molecular chains H$_{4}$ \textcolor{black}{and H$_{5}$}.


{\em Theory:} Consider a many-particle quantum system with at most pairwise interactions.  By Nakatsuji's theorem~\cite{Nakatsuji1976, Mazziotti1998b}, we have that there is a one-to-one mapping between the solutions of the Schr{\"o}dinger equation
\begin{equation}
\label{eq:SE}
\left ({\hat H} - E \right ) | \Psi \rangle = 0
\end{equation}
and the solutions of the contracted Schr{\"o}dinger equation (CSE)~\cite{Nakatsuji1976, Colmenero1993a, Nakatsuji1996, Yasuda1997, Mazziotti1998b, Mazziotti1999a, Coleman2000, Mazziotti2002d}
\begin{equation}
\label{eq:CSE}
\langle \Psi | {\hat a}^{\dagger}_{i} {\hat a}^{\dagger}_{j} {\hat a}^{ }_{l} {\hat a}^{ }_{k} \left ( {\hat H} - E \right ) | \Psi \rangle = 0.
\end{equation}
The proof of Nakatsuji's theorem follows from showing that the CSE implies the dispersion relation  $\langle \Psi | ( {\hat H} - E )^{2} | \Psi \rangle = 0$ which is true if and only if the wave function $| \Psi \rangle$ satisfies the Schr{\"o}dinger equation~\cite{Nakatsuji1976, Mazziotti1998b}.

Let us parameterize the Hamiltonian operator ${\hat H}$ in terms of a reference Hamiltonian operator ${\hat H}_{0}$, the perturbation operator ${\hat V}$, and the perturbation parameter $\lambda$
\begin{equation}
\label{eq:Hpt}
{\hat H}_{\lambda} = {\hat H}_{0} + \lambda {\hat V} .
\end{equation}
Differentiating the Schr{\"o}dinger equation yields
\begin{equation}
\label{eq:dSE}
\left ( {\hat H}_{\lambda} - E_{\lambda} \right ) \frac{d | \Psi_{\lambda} \rangle}{d\lambda} \textcolor{black}{+} \left ( {\hat V} - \frac{dE_{\lambda}}{d\lambda} \right ) | \Psi_{\lambda} \rangle = 0 .
\end{equation}
In Dalgarno-Lewis perturbation theory we express the derivative of $| \Psi_{\lambda} \rangle$ in terms of an operator ${\hat F}_{\lambda}$
 \begin{equation}
 \label{eq:F}
 \frac{d | \Psi_{\lambda} \rangle }{d\lambda}  = {\hat F}_{\lambda} | \Psi_{\lambda} \rangle
 \end{equation}
 Substituting Eq.~(\ref{eq:F}) into Eq.~(\ref{eq:dSE}) yields the equation for the  ${\hat F}_{\lambda}$ operator
 \begin{equation}
 \label{eq:DL}
 \left ( {\hat H}_{\lambda} - E_{\lambda} \right ) {\hat F}_{\lambda} | \Psi_{\lambda} \rangle  \textcolor{black}{+} \left ( {\hat V} - \frac{dE_{\lambda}}{d\lambda} \right ) | \Psi_{\lambda} \rangle = 0 .
 \end{equation}
 Using Rayleigh-Schr{\"o}dinger perturbation theory, we can formally express the ${\hat F}_{\lambda}$ as an $N$-body operator generated from the sum over all $N$-body eigenstate wave functions of the Hamiltonian $H_{\lambda}$.  \textcolor{black}{The operator ${\hat F}_{\lambda}$, however, is not uniquely defined because a convex set of operators will map one quantum state into another quantum state.  In the following discussion we exploit this non-uniqueness in combination with the CSE to prove that for a quantum system with any number of particles but at most pairwise interactions there always exists a two-body operator ${\hat F}_{\lambda}$ that satisfies Eq.~(\ref{eq:F}).}

 Consider differentiating the CSE in Eq.~(\ref{eq:CSE}) with respect to the perturbation parameter $\lambda$  and using the definition of ${\hat F}_{\lambda}$ in Eq.~(\ref{eq:F}) to obtain
 \begin{equation}
 \label{eq:CSEDL}
  \langle \Psi_{\lambda} | {\hat a}^{\dagger}_{i} {\hat a}^{\dagger}_{j} {\hat a}^{ }_{l} {\hat a}^{ }_{k} \left ( {\hat H}_{\lambda} - E_{\lambda} \right ) {\hat F}_{\lambda} + \left ( {\hat V} - \frac{dE_{\lambda}}{d\lambda} \right ) | \Psi_{\lambda} \rangle
 \end{equation}
 As discussed earlier, by Nakatsuji's theorem~\cite{Nakatsuji1976, Mazziotti1998b} there is a one-to-one mapping between the wave function solutions of the Schr{\"o}dinger equation and the CSE.   By continuity of the Schr{\"o}dinger equation and the CSE with respect to $\lambda$ the one-to-one mapping of Nakatsuji's theorem must hold for the differential forms of the Schr{\"o}dinger equation and the CSE in Eqs.~(\ref{eq:DL}) and~(\ref{eq:CSEDL}).  Hence, we have an extension of Nakatsuji's theorem to the differential Schr{\"o}dinger equation and CSE.  Furthermore, an ${\hat F}_{\lambda}$ operator solves the differential CSE in  Eq.~(\ref{eq:CSEDL}) if and only if it solves the differential Schr{\"o}dinger equation in Eq.~(\ref{eq:DL}).

 Despite the derivation of the contracted form of the Dalgarno-Lewis equation, we have not yet shown that the ${\hat F}_{\lambda}$ operator has a special form for Hamiltonians with only two-body interactions.  Consider the variational formulation of the Dalgarno-Lewis equation~\cite{1955a, Schwartz1959,Schiff1968}
 \begin{equation}
 \label{eq:vDL}
 {\min_{{\hat F}_{\lambda}} \Phi({\hat F}_{\lambda})},
 \end{equation}
 where
 \begin{eqnarray}
 \Phi({\hat F}_{\lambda}) & = & \langle \Psi_{\lambda} | \left ( {\hat V} - \frac{dE_{\lambda}}{d\lambda} \right ) {\hat F}_{\lambda} | \Psi_{\lambda} \rangle \\
                          & + & \langle \Psi_{\lambda} |  {\hat F}_{\lambda}^{\dagger} \left ( {\hat V} - \frac{dE_{\lambda}}{d\lambda} \right ) | \Psi_{\lambda} \rangle  \nonumber \\
                          & + & \langle \Psi_{\lambda} |  {\hat F}_{\lambda}^{\dagger} \left ( {\hat H}_{\lambda} - E_{\lambda} \right ) {\hat F}_{\lambda} | \Psi_{\lambda} \rangle  \nonumber
 \end{eqnarray}
 At the minimum we have
 \begin{equation}
 \label{eq:sc}
 \frac{\partial \Phi({\hat F}_{\lambda})}{\partial {\hat F}_{\lambda}} = 0 .
 \end{equation}
If we assume that ${\hat F}_{\lambda}$ is a two-body operator
\begin{equation}
\label{eq:F2}
{\hat F}_{\lambda} = \sum_{ijkl}{ ^{2} F_{\lambda}^{ij;kl} {\hat a}^{\dagger}_{i} {\hat a}^{\dagger}_{j} {\hat a}_{l} {\hat a}_{k} } ,
\end{equation}
then the stationary condition in Eq.~(\ref{eq:sc}) implies the differential CSE in Eq.~(\ref{eq:CSEDL}).  Because there is a one-to-one mapping between the solutions of Eq.~(\ref{eq:DL}) and Eq.~(\ref{eq:CSEDL}) by the extension of Nakatsuji's theorem to the differential limit, we have proved that the Dalgarno-Lewis equations---both  Eq.~(\ref{eq:DL}) and Eq.~(\ref{eq:CSEDL})---are satisfied by a Dalgarno-Lewis operator ${\hat F}_{\lambda}$ that is two-body.


\begin{table}

\caption{The hydrogen chain H$_{4}$ at each bond length $R$ (in angstroms) has a two-body Dalgarno-Lewis operator that satisfies both the Dalgarno-Lewis CSE (DL CSE) and the Dalgarno-Lewis (DL) equation.  The errors in the DL CSE and the DL are defined by the Frobenius norms of  their residuals.  The total energies are given in units of hartrees.}

\label{t:dl}

\begin{ruledtabular}
\begin{tabular}{cccccc}
     &  \multicolumn{1}{c}{Total}  & \multicolumn{2}{c}{Error in DL CSE} &  \multicolumn{2}{c}{Error in DL} \\
   \cline{3-4} \cline{5-6}
   \multicolumn{1}{c}{$R$}  &  \multicolumn{1}{c}{Energy}  & \multicolumn{1}{c}{1-body $F$} &  \multicolumn{1}{c}{2-body $F$} &  \multicolumn{1}{c}{1-body $F$} &  \multicolumn{1}{c}{2-body $F$} \\
   \hline
0.6	& -1.98120 	& 0.00461	& $2.51 \times 10^{-15}$  & 0.05113	& $2.47 \times 10^{-13}$ \\
1.0  &	-2.18097	    & 0.00581	& $2.28 \times 10^{-15}$  & 0.06430	& $4.25 \times 10^{-14}$ \\
1.4  & -2.04488	    & 0.00504	& $2.11 \times 10^{-16}$  & 0.03671	& $1.08 \times 10^{-15}$ \\
1.8	& -1.94221	    & 0.00218	& $2.96 \times 10^{-15}$  & 0.01358	& $1.06 \times 10^{-14}$ \\
2.2	& -1.90061	    & 0.00066	& $6.72 \times 10^{-16}$  & 0.00402	& $2.36 \times 10^{-15}$ \\
2.6	& -1.88828  	& 0.00019	& $5.15 \times 10^{-16}$  & 0.00115	& $1.79 \times 10^{-15}$
\end{tabular}
\end{ruledtabular}
\end{table}

Integration of the differential equation for the wave function in Eq.~(\ref{eq:F}) yields the wave function of the quantum system at $\lambda=1$.  If the $F_{\lambda}$ operators commute with each other, the solution can be expressed in closed form
\begin{equation}
\label{eq:PsiAppr}
| \Psi \rangle = e^{\int_{0}^{1}{{\hat F}_{\lambda} d \lambda} } | \Psi_{0} \rangle ,
\end{equation}
where the exponent is a two-body operator representing the integral of the two-body Dalgarno-Lewis operator.  In general, however, the $F_{\lambda}$ operators do not commute, and the solution cannot be expressed as a single two-body exponential transformation of the reference wave function.  Nonetheless, if we divide the integration over $\lambda$ into $M$ intervals on each of which $F_{\lambda}$ is nearly constant, we can express the wave function as
\begin{equation}
\label{eq:PsiEX}
| \Psi \rangle =  \prod_{k=1}^{M}{ e^{\epsilon {\hat F}_{k}} | \Psi_{0} \rangle } ,
\end{equation}
where ${\hat F}_{k}$ represents the integral of ${\hat F}_{\lambda}$ over $\lambda$ in the $k^{\rm th}$ interval.  Because the accuracy of the expansion can be arbitrarily improved by increasing $M$, we have derived an exact two-body exponential product expansion of the wave function.

While the expansion was derived by the author for the ground state in Ref.~\onlinecite{Mazziotti2004c}, the derivation from the perspective of Dalgarno-Lewis perturbation theory allows us to establish the expansion's \textcolor{black}{exactness for excited states as well as its rate of convergence.  Because Eq.~(\ref{eq:F}) is valid for both ground and excited states, we can generate an expansion of the wave function that is exact for not only the ground state but also the excited states.  To prove exactness, we select the $n^{\rm th}$ excited state of the initial reference Hamiltonian to be the initial reference wave function.  By propagating this excited state from the solution of the initial-value differential equation in Eq.~(\ref{eq:F}) as a function of the parameter $\lambda$, we generate an expansion that is exact for the $n^{\rm th}$ excited state, and hence, the expansion is exact for both ground and excited states.}

\textcolor{black}{Because the expansion solves Eq.~(\ref{eq:F}) without invoking the Taylor-series approximation of traditional perturbation theory, it} \textcolor{black}{is globally convergent to a stationary state of the Hamiltonian from an initial state of the reference Hamiltonian.  The rate of global convergence is inversely related to the number $M$ of product terms in the expansion required to achieved a given precision $\epsilon$ in the energy.  Each term in the product can be chosen to be a first-order step in the solution of Eq.~(\ref{eq:F}), and hence, $M$ is bounded from above by the number of first-order steps required to solve the differential equation to precision $\epsilon$.  Practically, $M$ can be made much smaller by optimizing the parameters in the expansion variationally (i.e. by a variational principle or solution of the CSE) rather than following the perturbative path of Eq.~(\ref{eq:F}), as shown in the results below.  Because each first-order transformation of the wave function corresponds to a second-order change in the energy, variational optimization of the energy with respect to the expansion's parameters exhibits local quadratic convergence---the energy converges quadratically in the vicinity of the solution.}

The two-body exponential product expansion can also be cast in other forms.  By expanding each exponential through first order, we have
\begin{equation}
\label{eq:PsiCI}
| \Psi \rangle = \prod_{k=1}^{M}{ (1+\epsilon {\hat F}_{k}) | \Psi_{0} \rangle } .
\end{equation}
 This form of the wave function was initially proposed by Nakatsuji~\cite{Nakatsuji2000}.  It remains exact and can also be obtained directly from Eq.~(\ref{eq:F}) by integrating  the differential equation by a first-order Euler method with $M$ steps.  \textcolor{black}{While Nakatsuji proved that this expansion is exact for the ground state, the present work shows that it is exact not only for the ground state but also excited states.  Furthermore,  from our analysis of Eq.~(\ref{eq:PsiEX}) it follows immediately that this expansion is also globally convergent at a rate at least as fast as the differential solution of Eq.~(\ref{eq:F}) with a quadratic local convergence of the energy.  Nonetheless, Eq.~(\ref{eq:PsiCI}) has a potential disadvantage relative to Eq.~(\ref{eq:PsiEX}).  Eq.~(\ref{eq:PsiCI}) is only size extensive in its energy upon convergence to the exact stationary state, but Eq.~(\ref{eq:PsiEX}) is size extensive in its energy upon truncation at any $M$ because the exponential generates the higher-order operators that are products of the lower-order operators.}  We can also express the differential equation for the wave function in Eq.~(\ref{eq:F}) as an integral equation
\begin{equation}
\label{eq:PsiFD}
| \Psi \rangle = \int_{0}^{1}{ {\hat F}_{\lambda} | \Psi_{\lambda} \rangle d \lambda} .
\end{equation}
Expansion of this integral equation in powers of ${\hat F}_{\lambda}$ generates an exact Feynman-like diagrammatic expansion in terms of two-body Dalgarno-Lewis operators.


\begin{table*}

\caption{For H$_{4}$ the error in the total energy, relative to full configuration interaction, is shown for several bond distances $R$ (in angstroms) from CSE(1) and CSE(2) as well as second-order many-body perturbation theory (MP2), coupled cluster with single-double excitations (CCSD), and CCSD with perturbative triple excitations (CCSD(T)).   The CSE(2) energies are exact to within the numerical convergence of the optimizer.  The energies and their errors are given in units of hartrees.}

\label{t:cse}

\begin{ruledtabular}
\begin{tabular}{cccccccc}
     &  \multicolumn{1}{c}{Total}  & \multicolumn{1}{c}{Correlation} & \multicolumn{5}{c}{Energy Error (hartrees)} \\
   \cline{4-8}
   \multicolumn{1}{c}{$R$}  &  \multicolumn{1}{c}{Energy}  & \multicolumn{1}{c}{Energy}  & \multicolumn{1}{c}{MP2} &  \multicolumn{1}{c}{CCSD} & \multicolumn{1}{c}{CCSD(T)} &  \multicolumn{1}{c}{CSE(1)} &  \multicolumn{1}{c}{CSE(2)} \\
   \hline
0.6	& -1.98120  &	-0.03071	  &  0.00869	&  ~0.000005  & -0.0000004 &  0.00012  &	 $4.80 \times 10^{-15}$ \\
1.0  &	-2.18097	  &   -0.06851  &  0.02700	&  ~0.000007	 & -0.0000048  &  0.00153  &  $2.00 \times 10^{-14}$ \\
1.4  & -2.04488	  &   -0.14234  &  0.06709	&  -0.000658	 & -0.0010788  &	0.01190  &	 $2.00 \times 10^{-14}$ \\
1.8  & -1.94221  &	-0.25768  &  0.12698	&  -0.009118	 & -0.0103678  &	0.04418  &	 $7.70 \times 10^{-13}$ \\
2.2	& -1.90061  &	-0.38321	  &  0.17058	&  -0.026423	 & -0.0290752  &	0.10168  &	 $9.99 \times 10^{-15}$ \\
2.6	& -1.88828  &	-0.48404  &  0.16432	&  -0.030854	 & -0.0353096  &	0.15575  & $1.64\times 10^{-13}$
\end{tabular}
\end{ruledtabular}
\end{table*}

{\em Results:} To demonstrate the validity and potential applicability of the theory, we examine the two-body expansion for the molecular hydrogen chains H$_{4}$ {\textcolor{black}{and H$_{5}$}.  Upon dissociation the chain undergoes a Mott metal-insulator transition~\cite{Suhai1994, Sinitskiy2010} with the insulator phase being strongly correlated due to spin entanglement.  The bonds are chosen to be equally spaced at a distance $R$, and we use a minimal Slater-type-orbital (STO-6G) basis set~\cite{Hehre1969}.  All calculations are performed with extensions to the Quantum Chemistry Package in the computer algebra system Maple.

According to the theory, for a Hamiltonian with only two-body interactions there exists a two-body Dalgarno-Lewis operator ${\hat F}$ that satisfies the Dalgarno-Lewis equation.  To demonstrate this result for the most general case in which the reference Hamiltonian is a two-body operator, we choose the reference Hamiltonian to be the full molecular Hamiltonian of H$_{4}$ rather than its one-body Hartree-Fock Hamiltonian and the perturbation to be the full molecular Hamiltonian minus the Hartree-Fock Hamiltonian.  In Table~I we show that the hydrogen chain H$_{4}$ at each bond length $R$ has a two-body Dalgarno-Lewis operator that satisfies the Dalgarno-Lewis CSE.  The Dalgarno-Lewis operator is computed from a linear least-squares solution of the Dalgarno-Lewis CSE.   For comparison we also show that the optimal one-body Dalgarno-Lewis operator has a nonzero minimum residual and hence, does not satisfy the Dalgarno-Lewis CSE.  In the table we also show for all bond lengths $R$ that the two-body ${\hat F}$ operator that satisfies the Dalgarno-Lewis CSE also solves the $N$-body Dalgarno-Lewis equation, which is consistent with the differential extension of Nakatsuji's theorem.


\begin{figure}
\begin{center}
\includegraphics[scale=0.38]{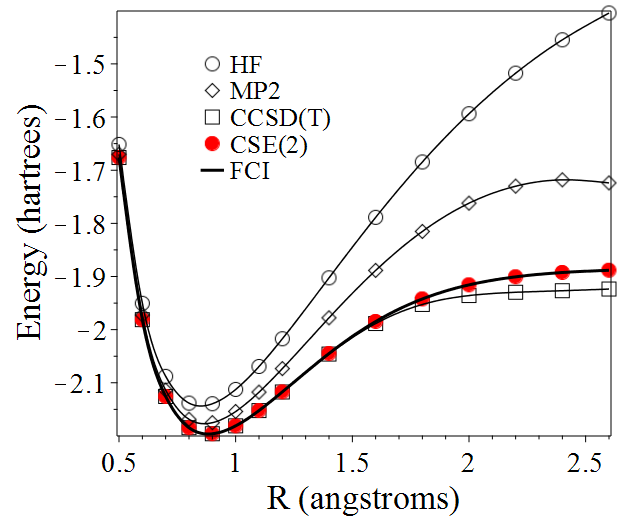}
\end{center}
\caption{The potential energy curves of H$_{4}$ from Hartree-Fock (HF), MP2, CCSD(T), CSE(2), and FCI are shown. The CSE(2) energies, shown by solid circles (red), are in exact agreement with the black line of FCI.}
\label{fig:cse}
\end{figure}

To illustrate the potential applicability of the theory, we use the two-body expansion of the wave function in Eq.~(\ref{eq:PsiCI}) to solve the CSE for H$_{4}$.  The use of $M=1$ and $M=2$ in Eq.~(\ref{eq:PsiCI}) to solve the CSE, we denote by CSE(1) and CSE(2), respectively.  \textcolor{black}{The expansion in Eq.~(\ref{eq:PsiCI}) is inserted directly into the CSE in Eq.~(\ref{eq:CSE}).}  The solution of  the resulting equation is performed by a least-squares minimization of the equation's residual using a limited-memory BFGS algorithm~\cite{Nocedal1980}.  Table~II shows the error in the total energy, relative to full configuration interaction (FCI), from CSE(1) and CSE(2) with comparisons to second-order many-body perturbation theory (MP2) as well as coupled cluster with single-double (CCSD) and perturbation triple (CCSD(T)) excitations.  Figure~1 presents the potential energy curves from Hartree-Fock (HF), MP2, CCSD(T), CSE(2), and FCI.   The CSE(2) energies, shown by red solid circles, are in agreement with the black line of FCI; in fact, the energy errors in Table~II show that the CSE(2) energies are exact  to within the numerical convergence of the optimizer.  Hence, the product of two of the Dalgarno-Lewis operators, we find, yields an exact wave function expansion for H$_{4}$ at all $R$.


\begin{table}

\caption{For H$_{4}$ the energies of the first 6 excited states from FCI and CSE(2) are reported in hartrees.}

\label{t:cse2}

\begin{ruledtabular}
\begin{tabular}{ccccccc}
   \multicolumn{1}{c}{State}  &  \multicolumn{1}{c}{$2$S$+1$}  & \multicolumn{1}{c}{FCI Energy}  & \multicolumn{1}{c}{CSE(2) Energy Error}   \\
   \hline
1	& 3 &	-1.954146208801  & $1.0 \times 10^{-15}$   \\
2  &	3 &  -1.862192277442  & $1.0 \times 10^{-15}$   \\
3  & 1 &  -1.824236275023  & $1.0 \times 10^{-15}$  \\
4  & 3 &	-1.759315766605  & $1.0 \times 10^{-14}$   \\
5	& 5 &	-1.702244608174  & $1.0 \times 10^{-14}$   \\
6	& 1 &	-1.584316227353  & $1.0 \times 10^{-15}$
\end{tabular}
\end{ruledtabular}
\end{table}

 \textcolor{black}{Using the CSE with the expansion in Eq.~(\ref{eq:PsiCI}) with $M=2$, we computed the first 6 excited states of H$_{4}$ at $R=1.4$~\AA\, shown in Table~\ref{t:cse2}.  The optimization procedure is the same as for the ground state with each state's reference wave function initialized to a different Hartree-Fock excited state.  As in the ground-state calculations the excited-state energies are exact to the precision limit of the floating-point arithmetic despite the strong correlation present at this stretched geometry.  The present calculations are very different from those of Nakatsuji who approximated the excited states from their response to the exact ground-state calculation~\cite{Nakatsuji2000}.  We also computed the lowest two doublet states of H$_{5}$.  The CSE(2) and FCI agree to machine precision in the states' energies of -2.538653212914 and -2.446397756519~hartrees.}

{\em Discussion and Conclusions:} The two-body expansions have connections to established methods.  The integral formulation of the expansion in Eq.~(\ref{eq:PsiFD}) shares the same mathematical structure as the generator of the Feynman diagrammatic expansion~\cite{Feynman1949}.  Nevertheless, there are significant differences between Feynman diagrams and the present method.  While Feynman diagrams are expressed in terms of the Hamiltonian operator as a function of time $t$ or frequency $\omega$ , the Feynman-like expansion of the generator in Eq.~(\ref{eq:PsiFD}) depends upon the two-body Dalgarno-Lewis operator as a function of the dimensionless perturbation parameter $\lambda$.  The exponential product formulation of the expansion in Eq.~(\ref{eq:PsiEX}) also has similarities to the coupled cluster expansion.  Both theories depend upon the exponential of operators.  However, while the exact coupled cluster expansion depends upon $p$-body excitation operators with $p$ ranging from one to the number of particles, the wave function expansion in Eq.~(\ref{eq:PsiEX}) depends upon only general 2-body operators.   The expansion in Eq.~(\ref{eq:PsiEX}) is a CSE ansatz rather than a conventional coupled cluster ansatz because the CSE acts as the stationary equation with respect to variations in each of the two-body operators.

Previous work established the two-body exponential expansion for the ground-state wave function~\cite{Mazziotti2004c}.  By connecting the expansion to a contracted formulation of Dalgarno-Lewis theory, the present work rigorously establishes that (i) the two-body operators generate a renormalized, globally convergent expansion,  (ii) the energy of the expansion converges quadratically near the solution, and (iii) the expansion is exact for both ground and excited states.  The two-body expansion offers a significant simplification of the many-particle wave function as well as the two-particle reduced density matrix that promises to provide more efficient and effective methods on both conventional and quantum computers~\cite{Lee2018, Xia2018, Grimsley2019, Evangelista2019, Smart2020} for the treatment of strong correlation.

\begin{acknowledgments}

D.A.M. gratefully acknowledges the Department of Energy, Office of Basic Energy Sciences Grant DE-SC0019215 and the U.S. National Science Foundation Grants No. CHE-1565638, No. CHE-2035876, and No. 296 DMR-2037783.

\end{acknowledgments}

\bibliography{DL}

\end{document}